\documentclass[pra,twocolumn,aps,10pt]{revtex4-1}

\usepackage{amsmath,amssymb}

\usepackage{graphicx,xcolor}

\usepackage[colorlinks=true]{hyperref}

\usepackage[utf8]{inputenc}
\usepackage[T1]{fontenc}

\newcommand{\ii}{\mathrm{i}}

\DeclareMathOperator{\tr}{tr}

\usepackage{bbm}

\begin{document}

\title{Immutable quantized transport in Floquet chains}

\author{Bastian H{\"o}ckendorf}
\author{Andreas Alvermann}
\author{Holger Fehske}
\affiliation{Institut f{\"u}r Physik, Universit{\"a}t Greifswald, Felix-Hausdorff-Stra{\ss}e~6, 17489 Greifswald, Germany}

\begin{abstract}

We show how quantized transport can be realized in Floquet chains through encapsulation of a chiral or helical shift.
The resulting transport is immutable rather than topological in the sense that it neither requires a band gap nor is affected by arbitrarily strong perturbations. Transport is still characterized by topological quantities
but encapsulation of the shift prevents topological phase transitions.
To explore immutable transport we introduce the concept of a shiftbox, explain the relevant topological quantities both for momentum-space dispersions and real-space transport, and study model systems of Floquet chains with strictly quantized chiral and helical transport.
Natural platforms for the experimental investigation of these scenarios are photonic Floquet chains constructed in waveguide arrays, as well as topolectrical or mechanical systems.
\end{abstract}

\maketitle

Topological features of physical systems are not only of fundamental theoretical interest~\cite{TKNN, haldane1988model, PhysRevB.78.195125_08, PhysRevX.8.031079_18, bergholtz2020exceptional}, but the robustness of topological phases
and of transport quantization against perturbations are of immediate experimental relevance~\cite{Klitzing, Rechtsman, Mukherjee, Maczewsky17, RevModPhys.91.015005_19}.
The relation between topology and transport is given by the bulk-boundary correspondence, which for the most important two-dimensional case can be summarized as follows:
A non-trivial topological phase of the two-dimensional bulk, characterized by a non-zero Chern number, is accompanied by topological boundary states that give rise to quantized chiral transport along the one-dimensional boundary. For a topological insulator with time-reversal symmetry (TRS), where the bulk topology is characterized by the $\mathbb Z_2$-valued Kane-Mele invariant, boundary states appear in pairs with opposite directionality and transport along the boundary is helical~\cite{KaneMelePRL05, Fu06, Konig2007, Hsieh08, HasanKane2010, RevModPhys.83.1057_11, GreifRostock}.

If we reduce the dimension from two to one, say in an attempt to separate the boundary transport from the bulk, we lose both topology and transport.
According to the mathematical classification of topological systems~\cite{RevModPhys.88.035005_16,PhysRevB.96.155118} a one-dimensional chain, either with TRS or without any symmetry, is in a topologically trivial state. One way out of this situation is provided by non-Hermitian Floquet chains that host non-trivial topological phases in dimension one~\cite{hckendorf2019nonhermitian, Fedorova2020, doi:10.1063/5.0036494}. However, transport is quantized only in the limit of strong non-Hermiticity.

At this point our present considerations come into play. Specifically, we will describe a strategy to implement strictly quantized chiral and helical transport in Hermitian Floquet chains. Our strategy might, at first, appear as an illegitimate trick, using the intellectual sleight of hand of redefining a problem in such a way that the proposed solution applies.
On further inspection, however, we simply employ a device (called a \emph{shiftbox}) that exploits intrinsic mechanisms of 
experimental platforms such as photonic waveguide arrays~\cite{SzameitJPB, RevModPhys.91.015006_19}, or topolectrical~\cite{Lee2018,2019arXiv190711562H} or mechanical~\cite{susstrunk2015,HuberMech16} systems. 

\begin{figure}
\includegraphics[width=\columnwidth]{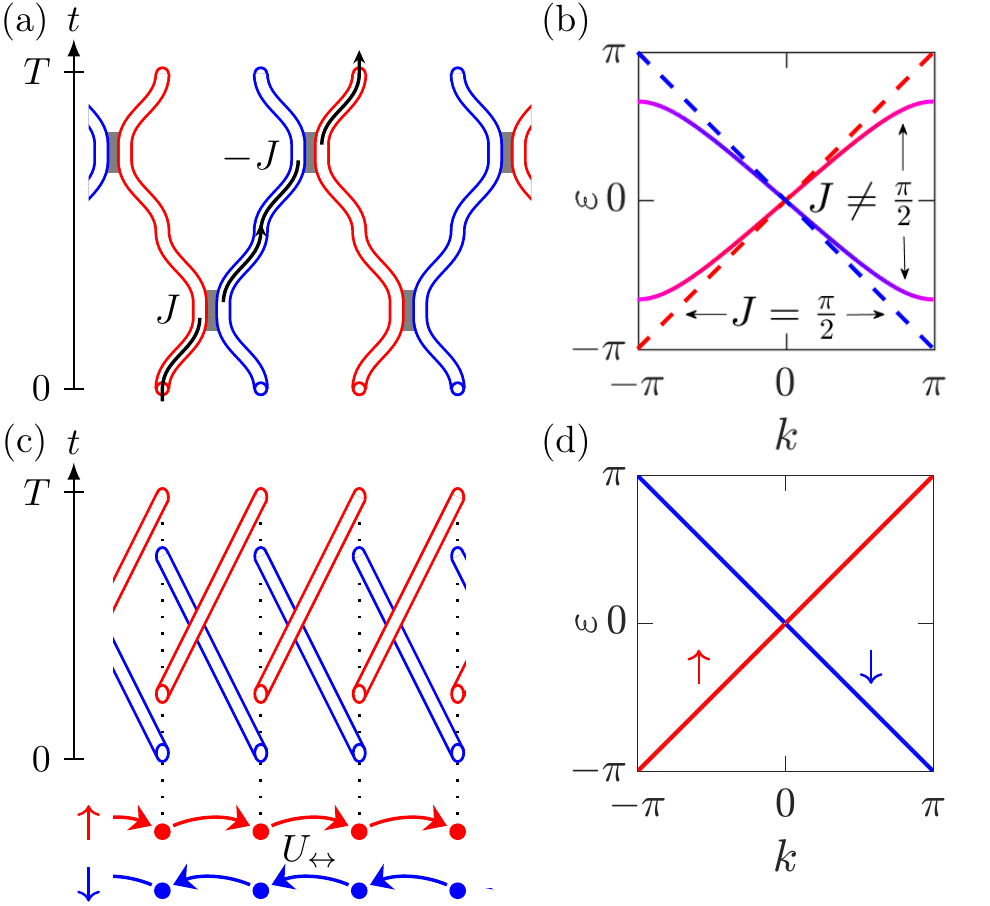}
\caption{ 
(a) Photonic waveguide implementation of one period of the Floquet chain from Ref.~\cite{PhysRevLett.118.105302}.
The vertical axis corresponds to time.
(b) Quasienergy dispersion $\varepsilon(k)$ in the general case ($J \ne \pi/2$, solid curves) and at perfect coupling ($J=\pi/2$, dashed lines).
(c) Photonic waveguide implementation of a helical shiftbox for the operator $U_\leftrightarrow$.
(d) Quasienergy dispersion $\varepsilon(k)$ associated with the helical shiftbox.
}
\label{Fig:1}
\end{figure}

All our considerations take place in the Floquet setting, with a time-periodic Hamiltonian $H(t) = H(t+T)$.
We are mainly interested in the Floquet propagator $U(T)$ that gives the stroboscopic time evolution after one (or multiple) periods~\cite{Hae97, Holthaus15}. It is obtained as the solution of the Schr\"odinger equation $\ii \partial_t U(t) = H(t) U(t)$ with initial condition $U(0)=\mathbbm 1$.

Floquet systems host `anomalous' topological phases not realized in static systems~\cite{KitagawaPRB, Rudner}.
The technical reason for the existence of such phases is the $2\pi$-periodicity
of the Floquet quasienergies $\varepsilon(k)$ that
allows for loops winding around the quasienergy Brillouin zone.
These loops carry topological information independently of a bulk,
which facilitates the deliberate engineering of anomalous Floquet phases without sacrificing their topological nature~\cite{HAF19, PhysRevLett.123.190403_19}.

Consider the Floquet chain in Fig.~\ref{Fig:1} (top row) that we borrow from Ref.~\cite{PhysRevLett.118.105302}.
We depict the Floquet chain as it could be implemented in a photonic waveguide array,
where time is represented by the vertical axis and the one spatial dimension by the horizontal axis.
Although our discussion is not restricted to this specific setting,
it is convenient to
consider a concrete experimental platform.

One period of the Floquet chain consists of two steps with alternating couplings between adjacent waveguides.
The Floquet Hamiltonian for the fundamental period $t \in [0,T)$ is
$H(t) = (2 J \delta/T) \sum_{n \in \mathbb Z} (|2n{+}\delta\rangle\langle 2n| + H.c.)$,
with $\delta=1$ for $0 \le t < T/2$ and $\delta=-1$ for $T/2 \le t < T$, and a coupling parameter $J$.

In a pseudo-spin notation
each individual coupling of two adjacent waveguides corresponds to a Hamiltonian of the form
 $J \sigma_x$, with the Pauli matrix $\sigma_{x}$.
The corresponding propagator
is given by the unitary matrix $\exp[-\ii J \sigma_x] =  (\cos J) \mathbbm 1 - (\ii \sin J) \sigma_x$.
Combining all couplings in the Floquet chain gives the Floquet propagator
\begin{multline}\label{Floq1}
 U_J(T) =  c^2   +\ii sc \sum_{n \in \mathbb Z} (-1)^n \Big (|n-1\rangle \langle n| 
-|n+1\rangle \langle n|\Big) \\
  +  s^2 \sum_{n \in \mathbb Z}\big( |2n+2\rangle \langle 2n|+|2n-1\rangle \langle 2n+1|\big) \, 
\end{multline}
for one period, with the abbrevations $c = \cos J$, $s = \sin J$.
From the eigenvalues $\xi$ of $U_J(T)$ we get the quasienergies $\varepsilon=\ii \log \xi$, which are normalized to $2 \pi$-periodicity.
The solid curves in Fig.~\ref{Fig:1}~(b) show the quasienergy dispersion $\varepsilon(k)$ at generic values of $J$.

After fine-tuning to a specific `perfect coupling' value (in our units, $J=\pi/2$)
each individual coupling in the Floquet chain results in
the full transfer of amplitude between two adjacent waveguides.
Therefore, an initial excitation follows the trajectory shown in Fig.~\ref{Fig:1}~(a),
and is shifted exactly by two waveguides in each period.
The Floquet propagator according to Eq.~\eqref{Floq1} is
\begin{equation}\label{eq:Floquet_prop_perf}
U_{\pi/2}(T) = \sum_{n \in \mathbb Z}\big( |2n+2\rangle \langle 2n|+|2n-1\rangle \langle 2n+1|\big)
\; .
\end{equation}
It describes quantized and, since the shift operates equally in both directions, helical transport.
The corresponding 
quasienergy dispersion is given by the two straight dashed lines $\varepsilon(k) = \pm k$ in Fig.~\ref{Fig:1}~(b).

We conclude that the Floquet chain supports quantized helical transport at---but only precisely at---the specific coupling $J=\pi/2$.
This effect is neither topological nor robust since any 
perturbation to $J \ne \pi/2$ opens a gap and destroys the quantized helical transport.
In other words, the present Floquet chain requires fine-tuning~\cite{PhysRevLett.118.105302}.

At this point we can make a
decisive observation:
For the realization of quantized transport in a Floquet system not the precise form of the Hamiltonian but only of the propagator $U(T)$ is relevant.
Now consider the waveguide array of Fig.~\ref{Fig:1}~(c), where waveguides in the top or bottom layer (red or blue in the figure) move one place to the right or left, respectively.
To denote the propagator implemented by this array, it is convenient to use pseudo-spin notation to distinguish the top ($|{\uparrow}\rangle$) and bottom ($|{\downarrow}\rangle$) waveguide layer.
Then, the propagator is the exact helical shift operator
\begin{equation}
U_\leftrightarrow=\sum_{n\in \mathbb Z} \big(|n+1,{\uparrow}\rangle\langle n,{\uparrow}|+ |n-1,{\downarrow}\rangle\langle n,{\downarrow}| \big)\;,
\label{eq:helical_shift}
\end{equation}
with the perfectly linear dispersion in Fig.~\ref{Fig:1}~(d).
Up to a change of notation, which accounts for the rearrangement of waveguides, 
this is precisely the propagator~\eqref{eq:Floquet_prop_perf} of the previous fine-tuned Floquet chain at perfect coupling.

We call the type of waveguide array in Fig.~\ref{Fig:1}~(c) a \emph{shiftbox}, because this describes what it is:
A device to encapsulate an exact shift propagator without having to refer to an underlying Hamiltonian or Floquet protocol. 

That a shiftbox works without reference to an underlying (Floquet) Hamiltonian is not a convenience but a necessity. Any Hamiltonian, not only the initial Hamiltonian leading to Eq.~\eqref{Floq1}, is subject to topological constraints that preclude robust (chiral or helical) transport in dimension one~\cite{RevModPhys.88.035005_16,PhysRevB.96.155118}.
Only the direct shiftbox implementation of a propagator without an underlying Hamiltonian can circumvent these constraints.
Shiftboxes are, therefore, unique to experimental platforms that allow for such implementations, e.g., via the setting of Fig.~\ref{Fig:1}~(c).

In contrast to the initial Floquet chain a shiftbox requires no fine-tuning.
There is no way to destroy the helical shift, or to open a gap in the linear dispersion, without breaking the shiftbox.
The effects of the shiftbox are entirely robust, and remain so in the experimental implementation. For example in a photonic system, it suffices to increase the distance between the waveguides to suppress any unwanted stray couplings.

Considered in isolation,
a shiftbox is a perfectly legitimate though not exceptionally interesting device:
It is engineered precisely to implement a helical shift and nothing more.
The helical shift according to Eq.~\eqref{eq:Floquet_prop_perf} is non-trivial but rigid by design.
Therefore, we will now ask whether shiftboxes still support helical transport if used as part of a general Floquet system.
In the photonic waveguide setting of Fig.~\ref{Fig:1}~(c) that would mean to allow for arbitrary couplings between adjacent photonic waveguides before and after the shiftbox array.

If we use a shiftbox as part of a general Floquet systems we can combine two types of dynamics:
The ``standard'' Floquet dynamics of continuous time evolution generated by a Hamiltonian $H(t)$ and given by the Schr\"odinger equation $\ii \partial_t U(t) = H(t) U(t)$,
and exceptional steps that directly specify the propagator via a shiftbox.

\begin{figure}
\includegraphics[width=\columnwidth]{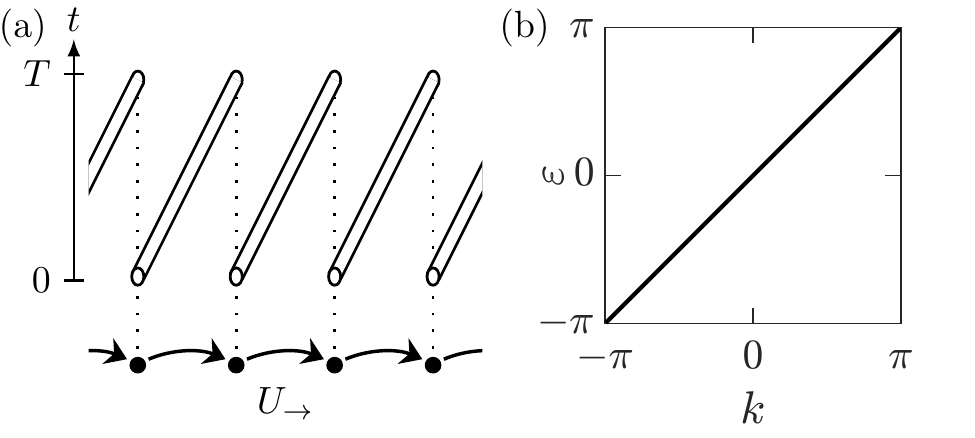}
\caption{
(a) Photonic waveguide implementation of a chiral shiftbox for the operator $U_\rightarrow$.
(d) Quasienergy dispersion $\varepsilon(k)$ associated with the chiral shiftbox.
}
\label{Fig:2}
\end{figure}

Before we explore helical transport from this angle, let us take a step back and consider chiral transport first.
A chiral shiftbox is depicted in Fig.~\ref{Fig:2}.
It implements the chiral shift (here, to the right)
\begin{equation}
U_\rightarrow=\sum_{n\in \mathbb Z} |n+1\rangle\langle n| \; .
\label{eq:chiral_shift}
\end{equation}
The chiral shiftbox is the top half of the helical shiftbox in Fig.~\ref{Fig:1}~(c),
which consists of two chiral shiftboxes with opposite directionality.

The chiral shift (to the right) corresponds to the perfectly linear quasienergy dispersion $\varepsilon(k) = k$,
that is, a single loop winding once around the quasienergy Brillouin zone.
Such a loop is immutable in the sense that it can be neither created nor destroyed in continuous time evolution generated by a Hamiltonian.
It is not difficult to see why: Loops are counted by the  winding number
\begin{equation}
\label{eq:winding}
\begin{aligned}
 W &= \frac{\ii}{2\pi} \int_{-\pi}^{\pi} \tr\big[U(T,k)^\dagger \partial_k U(T,k) \big] \, \mathrm{d}k \\
  &=  \frac{1}{2\pi} \sum_{n=1}^N \,  \int_{-\pi}^{\pi}   \partial_k \varepsilon_n(k) \, \mathrm{d}k \; ,
\end{aligned}
\end{equation}
an integer-valued topological quantity.
The trace in the first row of this expression runs over the $N$ sites of the unit cell,
the sum in the second row over the quasienergy bands $\varepsilon_1(k), \dots, \varepsilon_N(k)$.

Using the Schr\"odinger equation
for $U(t,k)$ and the cyclicity of the trace
one calculates the time-derivative  of the integrand in Eq.~\eqref{eq:winding} as
$\ii \partial_t \tr[U^\dagger \partial_k U] = \tr[U^\dagger (\partial_k H) U] = \tr[\partial_k H]$.
Integration $\int_{-\pi}^\pi \mathrm{d} k$ of this expression results in zero since $H(t,k)$ is periodic in $k$.
Therefore, the winding number $W$ is constant under continuous time evolution with a Hamiltonian $H(t)$.
On the other hand, introducing a chiral shiftbox into the dynamics changes $W$ by (plus or minus) one.

The constancy of $W$
implies that a chiral shift cannot be obtained from standard Floquet dynamics in any way, but requires an exceptional propagation step such as the shiftbox.
This is a stronger obstruction than for a helical shift~\eqref{eq:helical_shift}, which can be obtained from standard Floquet dynamics although only with fine-tuned parameters.
The obstruction corresponds to the fact that one-dimensional systems classify as topologically trivial.
Conversely, it implies the immutability of the chiral shift, which cannot be destroyed by standard Floquet dynamics.

\begin{figure}
\includegraphics[width=\columnwidth]{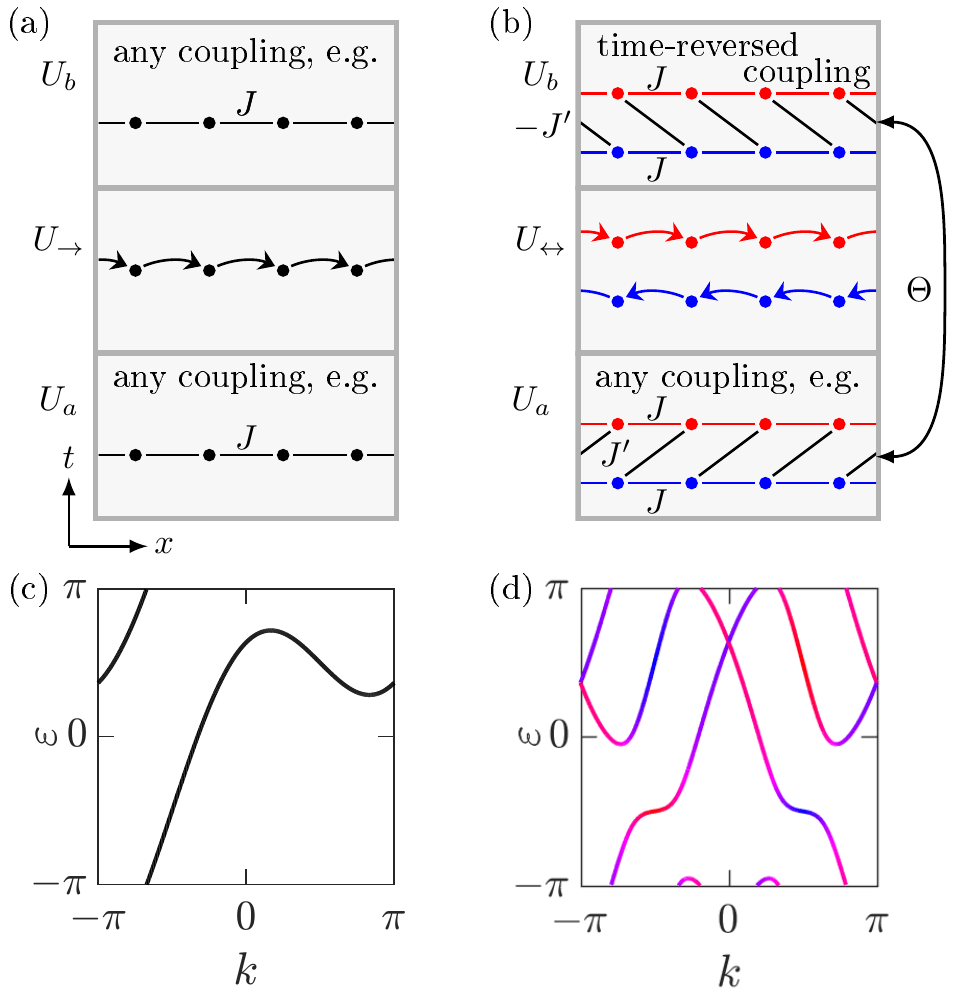}
\caption{(a) Chiral shiftbox sandwiched between standard Floquet dynamics. 
(b) Helical shiftbox sandwiched between standard Floquet dynamics with fermionic TRS.
(c) Quasienergy dispersion $\varepsilon(k)$ for the sandwiched chiral shiftbox with $J=1$ (see text for definitions).
(d) Quasienergy dispersion $\varepsilon(k)$ for the sandwiched helical shiftbox with $J=J'=1$.
\label{Fig:3}
}
\end{figure}

Fig.~\ref{Fig:3}~(a) depicts a general Floquet system with a chiral shiftbox sandwiched between standard Floquet dynamics.
The Floquet propagator has the form $U(T) = U_b U_\rightarrow U_a$,
where both $U_a, U_b$ obey a Schr\"odinger equation but the chiral shift $U_\rightarrow$ does not.
The precise form of the Hamiltonians $H_{a,b}(t)$ generating $U_{a,b}$ is not relevant.
By the above arguments, the winding number associated with $U(T)$ is $W=1$ for any choice of $H_{a,b}$.

For the quasienergy dispersion in Fig.~\ref{Fig:3}~(c) we choose the parametrization $H_{a,b}(t) \equiv H_J$
with $H_J = (3J/T) \sum_{n \in \mathbb Z} ( |n+1\rangle\langle n| + H.c.)$, a simple nearest-neighbor coupling between adjacent waveguides as it might occur in the experiment. Although the dispersion is strongly deformed in comparison to Fig.~\ref{Fig:2}~(b) one recognizes the single loop with $W=1$ generated by the chiral shiftbox.

The winding number $W$ has a strict relation to transport.
For a translational-invariant chain, we may measure transport by the propagation distance after one Floquet period, which is given by
\begin{equation}
\bar C = \frac{1}{N} \sum_{n=1}^N \langle n| U^{\dagger}(T) (\hat x-n) U(T) |n \rangle \; .
\label{eq:transport}
\end{equation}
In this expression,
$\hat x=\sum_{n\in \mathbb Z} n\, |n\rangle \langle n|$ is the position operator,
and we sum over the $N$ sites of the unit cell.
A Fourier transformation between position and momentum space gives the relation
\begin{equation}
\bar C=W
\label{topeqtrans}
\end{equation}
between transport and topology~\cite{hckendorf2019nonhermitian}.
This relation implies that transport in a one-dimensional chain is strictly quantized and, by the constancy of $W$, immutable.

Restricting ourselves to standard Floquet dynamics, this statement is empty as we always have $\bar C = W = 0$.
A one-dimensional chain does not support unidirectional transport.
Leaving the realm of standard Floquet dynamics, $\bar C = W \ne 0$ becomes possible through use of a chiral shiftbox.
In this way, the chain supports unidirectional transport, where wave packets are translated by an integer number of sites per period. Now, the strict quantization of transport follows from the identification of transport with topology in Eq.~\eqref{topeqtrans}.

In our interpretation of this result,
the chiral transport arising from a shiftbox is not topological in the sense that its quantization is protected by a topological invariant. A topological invariant is a quantity that 
may change its value in a topological phase transition, while the topological quantity $W$ occuring here is invariably constant under standard Floquet dynamics.
The constancy of $W$ strictly separates phases
with $\bar C=W=0$ and $\bar C=W \ne 0$. A transition between the phases cannot occur unless we modify the system externally by adding or removing the shiftbox.
Therefore, we designate the quantized transport \emph{immutable} rather than topological.
This designation agrees with the classification of chains without symmetries as topologically trivial~\cite{RevModPhys.88.035005_16,PhysRevB.96.155118}.

Two chiral shiftboxes can be combined with the same or opposite directionality. With the same directionality the winding number is $W=\pm 2$, and immutable quantized transport exists.
With opposite directionality we get the helical shiftbox in Fig.~\ref{Fig:1}~(c), with $W=1-1=0$.
Equivalently, we compute $\bar C=1-1=0$ for the bidirectional transport implemented by the helical shift~\eqref{eq:helical_shift}.

That $\bar C=W=0$ for the helical shiftbox implies that bidirectional transport is not protected (or immutable) without an additional constraint.
Just as for the two-dimensional topological $\mathbb Z_2$ insulators~\cite{KaneMelePRL05} this constraint is provided by fermionic TRS.

Fermionic TRS imposes the condition~\cite{PhysRevB.96.155118}
\begin{equation}
\Theta U(T) \Theta^{-1}= U^\dagger(T) 
\label{eq:TRS}
\end{equation}
on the Floquet propagator,
with an anti-unitary symmetry operator $\Theta$ such that $\Theta^2=-1$.
In the pseudo-spin notation of Eq.~\eqref{eq:helical_shift} we choose $\Theta =\sigma_y \mathcal K$,
with the Pauli matrix $\sigma_y$ and the operator of complex conjugation $\mathcal K$.
Note that in contrast to a standard Floquet system the TRS condition is imposed only on the propagator $U(T)$ after one period, not on the function $t \mapsto U(t)$.

Now consider a general Floquet system with a helical shiftbox sandwiched between standard Floquet dynamics, as depicted in Fig.~\ref{Fig:3}~(b).
The Floquet propagator has the form $U(T) = U_b U_\leftrightarrow U_a$, in analogy to the chiral case.
Again, the precise form of the Hamiltonians $H_{a,b}(t)$ generating $U_{a,b}$ is not relevant.
Here, we choose
\begin{multline}
H_{a,b}(t) =  H_J \otimes \mathbbm 1_\mathrm{spin}
\\
 + (3 J_{a,b}/T) \sum_{n\in \mathbb Z} \big(|n+\delta_{a,b},{\uparrow} \rangle \langle n,{\downarrow}| + H.c.\big)
\label{eq:HAB}
\end{multline}
with the nearest-neighbor coupling $H_J$ used before, and a second term with  $\delta_a=1$, $\delta_b=-1$
that couples the pseudo-spin components as depicted in the figure.

Without fermionic TRS, the $U_{a,b}$ are arbitrary and the helical dispersion is not protected.
With fermionic TRS, the $U_{a,b}$ are related by $U_b^\dagger = \sigma_y U_a^* \sigma_y$. In the parametrization of Eq.~\eqref{eq:HAB} we have to set $J_a = - J_b = J'$.
Now, Kramers degeneracy protects the crossings at $k=0, \pi$ in the dispersion shown in Fig.~\ref{Fig:3}~(d).
Although the dispersion is strongly deformed in comparison to Fig.~\ref{Fig:1}~(d) one clearly recognizes the two helical loops.

Note that the original Floquet protocol in Fig.~\ref{Fig:1}~(a)
is not protected by TRS, which is why the helical dispersion requires fine-tuning to the special value $J=\pi/2$.
The lack of TRS is not an oversight. Instead, the classification of one-dimensional topological phases implies that
a helical dispersion cannot be generated through continuous time evolution with any TR symmetric Hamiltonian. Otherwise, a TR symmetric topological phase would exist in disagreement with the negative result from the classification.
In dimension one, fermionic TRS gives rise to new robust phases only with a helical shiftbox.

As for the chiral shift~\eqref{eq:chiral_shift} we seek a relation between topology and transport.
To count helical loops we can consider the sum for the winding number in Eq.~\eqref{eq:winding} and note that, because of TRS, quasienergy loops appear in pairs with $\varepsilon_{n'}(k) = \varepsilon_n(-k)$ (as directly seen in Fig.~\ref{Fig:3}~(d)).
Summing over only one representative of each pair we get a $\mathbb Z_2$-valued winding number $W_\Theta$~\cite{doi:10.1063/5.0036494},
in close analogy to the construction for topological insulators~\cite{PhysRevB.75.121306_07,Carpentier,Nathan15,PhysRevB.96.195303_17,HockendorfPRB}.
$W_\Theta$ distinguishes between trivial phases with an even ($W_\Theta=0$)
and non-trivial phases with an odd number ($W_\Theta=1$ in Fig.~\ref{Fig:3}~(d)) of pairs of helical loops.

Note that for $J' \ne 0$ the two pseudo-spin components in our example are fully coupled.
In difference to Fig.~\ref{Fig:1}~(d) the helical loops in Fig.~\ref{Fig:3}~(d) are not associated with a definite spin orientation. Instead, the orientation depends on momentum $k$. This is the general behavior to be accounted for during evaluation of $W_\Theta$ or measurement of helical transport. To capture this behavior we choose the second term in the Hamiltonian~\eqref{eq:HAB} as given, instead of a simpler term $J_{a,b} \, \sigma_x$ that would not lead to momentum-dependent pseudo-spin mixing.

\begin{figure}
\includegraphics[width=\columnwidth]{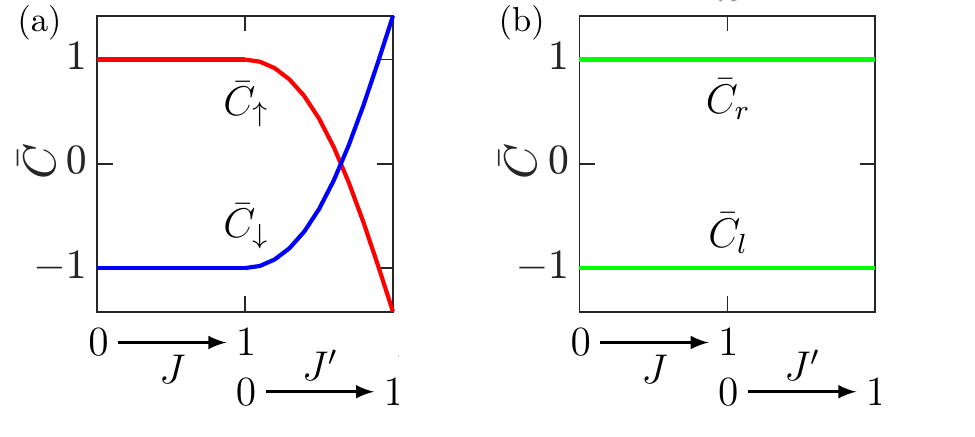}
\caption{Bidirectional transport with the helical shiftbox.
(a) Transport measured per pseudo-spin direction ($\bar C_\uparrow$, $\bar C_\downarrow$) is quantized only for uncoupled pseudo-spin components, that is for $J'=0$ (see text).
(b) Transport measured for Wannier functions ($\bar C_r$, $\bar C_l$) is strictly quantized for all parameters.
\label{Fig:4}
}
\end{figure}

Measuring helical transport requires us to separate the two transport directions.
The quantity $\bar C$ in Eq.~\eqref{eq:transport} sums indiscriminately over wavefunctions propagating to the left and right,
and thus is zero for bidirectional helical transport (in agreement with $\bar C=W=0$).

For $J'=0$ in Eq.~\eqref{eq:HAB} a right or left propagating wave function has fixed pseudo-spin $|{\uparrow}\rangle$ or $|{\downarrow}\rangle$.
Thus we can measure transport $\bar C_\uparrow=1$, $\bar C_\downarrow=-1$ separately
for each spin component (see Fig.~\ref{Fig:4}~(a)).
This is essentially the procedure for chiral transport applied twice.
But coupling between the pseudo-spin components for $J' \ne 0$  destroys the strict right vs. left separation,
and thus the individual quantization of $\bar C_{\uparrow,\downarrow}$. Only the sum $\bar C=\bar C_\uparrow + \bar C_\downarrow$ is quantized, but trivially equal to zero.

To measure helical transport in the general case, with coupling between the pseudo-spin components, we construct Wannier functions~\cite{Wannier37} from the Bloch functions of either one of the two quasienergy loops of the helical dispersion.
The Wannier functions appear in pairs $|\psi_r\rangle$, $|\psi_l\rangle$, and are mapped onto each other by the TRS operator $\Theta$~\cite{doi:10.1063/5.0036494}.
For $J' \ne 0$ each Wannier function is a superposition of all pseudo-spin directions.

We can now evaluate $\bar C$ separately for each Wannier function $|\psi_r\rangle$, $|\psi_l\rangle$.
Since a Wannier function is associated with a definite propagation direction,
$\bar C_{r,l}$ is strictly quantized even for $J'\ne 0$ (see Fig.~\ref{Fig:4}~(b)).
The quantization holds for any choice of the standard Floquet dynamics $U_{a,b}$,
as long as fermionic TRS is satisfied.
We can follow our earlier interpretation and recognize the strict quantization as a signature of immutable (now bidirectional) transport.
It requires a helical shiftbox together with fermionic TRS of the standard Floquet dynamics.

In conclusion, shiftboxes are devices that overcome a fundamental restriction of standard Floquet dynamics which does not support robust chiral or helical transport in dimension one.
Shiftboxes work by directly implementing strictly quantized unidirectional or bidirectional transport outside of the continuous time evolution governed by a Hamiltonian.
They can be natural components of experimental platforms that allow for explicit manipulation of particle or excitation trajectories.
In addition to our primary example of photonic waveguide arrays, where one can arrange
the waveguides in the desired shiftbox pattern, this includes topolectrical circuits or mechanical systems where one can freely arrange the electrical wires or mechanical springs. Shiftboxes might also work for lattices of trapped atoms if one directly manipulates  atomic positions via the trapping potential.

Since the transport generated by shiftboxes is immutable in the sense that it cannot be destroyed by standard Floquet dynamics,
shiftboxes can be combined with any type of continuous time evolution without giving up transport quantization.
This allows for deliberate engineering of wave function dynamics on top of the quantized transport.
A possible application might be the controlled transport of quantum information, encoded in the wave function amplitude, over multiple Floquet cycles.
Further exploration of the theoretical and experimental potential of this situation should prove worthwhile.

\end{document}